\documentclass[twocolumn,eqsecnum,showkeys,showpacs,nofootinbib,aps,epsfig]{revtex4}

\usepackage{graphicx}
\def\bea{\begin{eqnarray}}
\def\eea{\end{eqnarray}}

\newcommand{\nn}{\nonumber}

\def\beq{\begin{equation}}
\def\eeq{\end{equation}}

\def\pa{\partial}

\newbox\pippobox

\begin{document}
\input epsf
\title{Higher dimensional cosmological model with a phantom field}
\author{Soon-Tae Hong} \email{soonhong@ewha.ac.kr}
\affiliation{Department of Science Education and Research Institute
for Basic Sciences,\\ Ewha Womans University, Seoul 120-750 Korea}
\author{Joohan Lee}
\email{joohan@kerr.uos.ac.kr}
\affiliation{Department of Physics,
University of Seoul, Seoul 130-743 Korea}
\author{Tae Hoon Lee}\email{thlee@ssu.ac.kr}
\affiliation{Department of Physics and Institute of Natural
Sciences,\\ Soongsil University, Seoul 156-743 Korea}
\author{Phillial Oh}
\email{ploh@newton.skku.ac.kr}\affiliation{Department of Physics and
Institute of Basic Science, Sungkyunkwan University, Suwon 440-746
Korea}
%\date{January 22, 2003}%
\date{\today}
\begin{abstract}
We consider a higher dimensional gravity theory with a negative
kinetic energy scalar field and a cosmological constant.  We find
that the theory admits an exact cosmological solution for the scale
factor of our universe.  It has the feature that the universe
undergoes a continuous transition from deceleration to acceleration
at some finite time. This transition time can be interpreted as that
of recent acceleration of our universe.
\end{abstract}
%\preprint{hep-th/yymmnn}
\maketitle

%%%%%%%%%%%%%%%%%%%%%%%%%%%%%%%%%%%%%%%%%%%%%%%%%%%%%%%%
%\section{Introduction}
%\setcounter{equation}{0}
%\renewcommand{\theequation}{\arabic{section}.\arabic{equation}}
%%%%%%%%%%%%%%%%%%%%%%%%%%%%%%%%%%%%%%%%%%%%%%%%%%%%%%%%

Since Edwin Hubble discovered the expansion of our universe, the
inflationary big bang cosmology has been developed into a precision
science by recent cosmological observations including supernova
data~\cite{ob} and measurements of cosmic microwave background
radiation~\cite{ob2}. They suggest that our universe is made up of
about 4 percent ordinary matter, about 22 percent dark matter, and
about 74 percent dark energy. These observations triggered an
explosion of recent interests in the origin of dark
energy~\cite{dark}. There are several approaches to understanding
the dark energy such as cosmological constant,
quintessence~\cite{cald}, k-essence~\cite{a, c, m, nojiri} and
phantom~\cite{phantom}. One of the simple ways to explain it is
through an introduction of the cosmological constant, which leads to
an exponential expansion of the scale factor of the universe via
$a(t)\sim e^{\sqrt{\Lambda/3}~t}$ with the four-dimensional
cosmological constant $\Lambda$. Even though this simple form
adequately describes current acceleration with a small value of
$\Lambda$, it cannot say anything about important issues including
coincidence problem. If we could obtain a cosmological solution of
the form $a(t)\sim e^{F(t)}$ with $F(t)$ having the property that
the second time derivative of $a(t)$ vanishes at some finite time,
we may be able to interpret that time as the transition time from
deceleration to acceleration of the universe.

We appeal to higher dimensions because it turns out that the
4-dimensional version of our model (See Eq. (2).) does not admit
aforementioned type of cosmological solution.
In this paper, we show
that this type of solution is viable in higher
dimensional~\cite{freu, comment} phantom cosmology. We find that the
theory allows a cosmological solution of the scale factor of our
universe in which the standard exponential acceleration is modified
by higher dimensional phantom contribution. The solution exhibits an
interesting property that the universe undergoes a continuous
transition from deceleration to acceleration at some finite time.
This time could be interpreted as beginning of recent acceleration
of our universe.

A phantom energy component is characterized by the equation of state
parameter $\omega$ less than $-1$. Since Caldwell~\cite{phantom}
pointed out that the current observations do not rule out the
possibility of such an energy component and the universe may end its
existence in a finite time by reaching a singularity known as the
``big rip" according to some simple models, phantom cosmology has
been widely studied. In particular, it was suggested by
Scherrer~\cite{sche} that the finite lifetime of the universe may
provide an answer to the cosmic coincidence problem. The simplest
phantom model is constructed in terms of a scalar field with
negative kinetic energy.

Motivated by the recent interest of the ten-dimensional superstring
theories \cite{string} as the unified theories of fundamental
interactions and by the finding \cite{dabrowski} that there exist
interesting dualities between phantom and ordinary matter models
which are similar to dualities in superstring cosmologies
\cite{strcos, strcos2}, we extend this phantom model to a
ten-dimensional spacetime with a (ten-dimensional) cosmological
constant $\bar{\Lambda}(>0)$. In this model the equation of state is
given by \beq
\omega\equiv{p\over\rho}=\frac{-{1\over2}\dot{\sigma}^2-V(\sigma)-\bar{\Lambda}}
{-{1\over2}\dot\sigma^2+V(\sigma)+\bar{\Lambda}},\label{omega}\eeq
where $\sigma$ is the phantom field. If we insist on the positivity
of the energy, then it implies that $\omega\le-1$, with the equality
holding when the scalar field is constant. Unlike in the
four-dimensional case, $\omega\le-1$ does not guarantee the
acceleration of our universe due to the presence of scale factor of
extra dimensions.

Let us consider the action of the form \beq S=\int
d^{10}x\sqrt{-g}\left(-\frac{1}{2}R+\frac{1}{2}g^{MN}\pa_{M}\sigma\pa_{N}\sigma-V(\sigma)-\bar{\Lambda}\right).
\eeq The equations of motion are given by \bea
R_{MN}-\frac{1}{2}g_{MN}R-g_{MN}\bar{\Lambda}&=&-T_{MN}\nn\\
\frac{1}{\sqrt{-g}}\pa_{M}(\sqrt{-g}~g^{MN}\pa_{N}\sigma)+\frac{\pa
V}{\pa \sigma}&=&0. \eea Let us assume the metric of the form \beq
g_{MN}=\left(
\begin{array}{ccc}
-1 &0 &0\\
0  &a^{2}(t) &0\\
0 &0 &b^{2}(t)
\end{array}
\right), \eeq where $a(t)$ is the scale factor of our
three-dimensional universe and  $b(t)$ is the scale factor of the
extra six dimensions. We assume that the internal space is given by
six-dimensional torus~\cite{torus} and the potential vanishes,
$V(\sigma)=0$.

Exploiting the above equations, we have four differential equations
\bea
3\frac{\ddot{a}}{a}+6\frac{\ddot{b}}{b}&=&\frac{1}{4}\bar{\Lambda}+\dot{\sigma}^{2},\nn\\
\frac{\ddot{a}}{a}+2\frac{\dot{a}^{2}}{a^{2}}+6\frac{\dot{a}}{a}\cdot\frac{\dot{b}}{b}&=&\frac{1}{4}\bar{\Lambda},\nn\\
\frac{\ddot{b}}{b}+5\frac{\dot{b}^{2}}{b^{2}}+3\frac{\dot{a}}{a}\cdot\frac{\dot{b}}{b}&=&\frac{1}{4}\bar{\Lambda},\nn\\
\ddot{\sigma}+\left(3\frac{\dot{a}}{a}+6\frac{\dot{b}}{b}\right)\dot{\sigma}&=&0,\label{eoms}\eea
where the overdots denote the derivatives with respect to time.

Now, we define  variables $\alpha$ and $\beta$ as \beq
\alpha=H-h,~~~\beta=H+h.\eeq with $H=\dot{a}/a$ and $h=\dot{b}/b$.
After some manipulations, we rewrite (\ref{eoms}) in terms of
$\alpha$ and $\beta$ as \bea
3\alpha^{2}-\frac{4}{3}\frac{d}{dt}\left(\frac{\dot{\alpha}}{\alpha}\right)&=&\frac{3}{2}\dot{\sigma}^{2},\nn\\
\frac{3}{2}(3\beta-\alpha) + \left(\frac{\dot{\alpha}}{\alpha}\right)&=&0,\nn\\
\ddot{\sigma}-\left(\frac{\dot{\alpha}}{\alpha}\right)\dot{\sigma}&=&0.\label{pqeqs}\eea

For very late time ($t\rightarrow \infty$) we assume that
$\dot{\sigma}\sim 0$ asymptotically so that the solution approaches
the maximally symmetric ten-dimensional spacetime with cosmological
constant $\bar{\Lambda}$, $a\propto
e^{\frac{1}{6}\sqrt{\bar{\Lambda}}t}$ and $b\propto
e^{\frac{1}{6}\sqrt{\bar{\Lambda}}t}$. In this case one can show
that $\frac{\dot{\alpha}}{\alpha}\sim{ constant}$. In order to find
an exact solution\footnote{In recent time ($t<<\infty$), both the
phantom field and extra dimensions are necessary to have a
non-trivial solution of the form Eq. (12).} we take the following
ansatz \beq \frac{\dot{\alpha}}{\alpha}=\nu={ constant},
\label{condi}\eeq which is consistent with the asymptotic behavior.
From the physical point of view, this ansatz corresponds to the case
in which the volume of the universe as a ten-dimensional spacetime
increases at a constant rate.

With the above ansatz the solutions to (\ref{pqeqs}) are
then given by \bea \alpha&=&\alpha_{0}e^{\nu t},\nn\\
\beta&=&-\frac{2}{9}\nu+\frac{1}{3}\alpha_{0}e^{\nu t},\nn\\
\sigma&=&-\frac{\sqrt{2}}{\nu}\alpha_{0}e^{\nu t},\label{pqs} \eea
where $\alpha_{0}$ is the initial value of $\alpha$ at $t=0$ and
$\nu$ is given by \beq \nu=-\frac{3}{2}\sqrt{\bar{\Lambda}}.\eeq We
define $\alpha_{0}$ in terms of a dimensionless variable $n$ \beq
\alpha_{0}=n\sqrt{\bar{\Lambda}}.\eeq We then arrive at the solutions
to the differential equations (\ref{eoms}) as follows
\bea a&=&a_{0}~e^{\frac{1}{6}\sqrt{\bar{\Lambda}}t}e^{-\frac{4n}{9}(e^{-\frac{3}{2}\sqrt{\bar{\Lambda}}t}-1)},\nn\\
b&=&b_{0}~e^{\frac{1}{6}\sqrt{\bar{\Lambda}}t}e^{\frac{2n}{9}(e^{-\frac{3}{2}\sqrt{\bar{\Lambda}}t}-1)},\nn\\
\sigma&=&\sigma_{0}~e^{-\frac{3}{2}\sqrt{\bar{\Lambda}}t},\label{solns}\eea
where $a_{0}$ and $b_{0}$ are the initial values of $a$ and $b$ at
$t=0$, respectively, and  \beq
\sigma_{0}=\frac{2\sqrt{2}n}{3}.\label{sigma0}\eeq Moreover,
$\omega$ in (\ref{omega}) is then given by \beq
\omega=\frac{n^{2}e^{-3\sqrt{\bar{\Lambda}}t}+1}{n^{2}e^{-3\sqrt{\bar{\Lambda}}t}-1}.\label{omega2}\eeq
Here one notes that exploiting the relation
$\bar{\Lambda}=12\Lambda$ between the ten-dimensional cosmological
constant $\bar{\Lambda}$ and the four-dimensional one $\Lambda$,
in the vanishing phantom field limit with $n=0$, the solution for
$a$ in (\ref{solns}) is reduced to the standard form $a(t)\sim
e^{\sqrt{\Lambda/3}~t}$ of the exponential expansion of the scale
factor of the universe. In order to investigate the transition
time $t_{tr}$ from deceleration to acceleration, we consider the
differential equation \beq
\frac{\ddot{a}}{a}=\left(\frac{\dot{a}}{a}\right)^{2}-n\bar{\Lambda}
e^{-\frac{3}{2}\sqrt{\bar{\Lambda}}t}=0,\eeq whose solution is
given by \beq t_{tr}= \frac{2}{3\sqrt{\bar{\Lambda}}}\left(\ln
n-\ln \frac{7-\sqrt{45}}{8}\right).\label{tt} \eeq

In order to analyze the solutions (\ref{solns}) further near the
present time, we define a dimensionless variable $x$ as \beq
t=xt_{*}\eeq where the age of the universe is given by
$t_{*}=13.58~{\rm Gyr}=4.28\times 10^{17}$ sec~\cite{hartnett}.
The vacuum energy density is given by $\rho_{\Lambda}=10^{-8}~{\rm
erg}\cdot{\rm cm}^{-3}$~\cite{carroll} to yield $\Lambda=8\pi
G\rho_{\Lambda}=2.07\times 10^{-56}~{\rm cm}^{-2}$.  Taking the
ansatz that the transition time from deceleration to acceleration
in (\ref{tt}) is equal to $t_{tr}=5.04$ Gyr$=0.37
t_{*}$~\cite{hartnett}, we fix the value of $n$ to be $n=1.277$.
We then have $a$ and $b$ in terms of $a_{*}$ and $b_{*}$ as
follows \bea
a&=&a_{*}e^{1.07(x-1)-0.568(e^{-9.63x}-e^{-9.63})},\nn\\
b&=&b_{*}e^{1.07(x-1)+0.284(e^{-9.63x}-e^{-9.63})},\label{aabb}\eea
where $a_{*}=4.60\times 10^{10}$ light-years $=4.35\times 10^{28}$
cm~\cite{davis05} is the size of the our present universe and
$b_{*}$ is the size of the present extra dimensional space.

%%%%%%%%%%%%%%%%%%%%%%%%%%%%%%%%%%%%%%%%%%%%%%%%%%%%%%%%%%%%%%%%%%%%%%%%%%%%%
\begin{figure}
\begin{center}
\includegraphics[width=7cm]{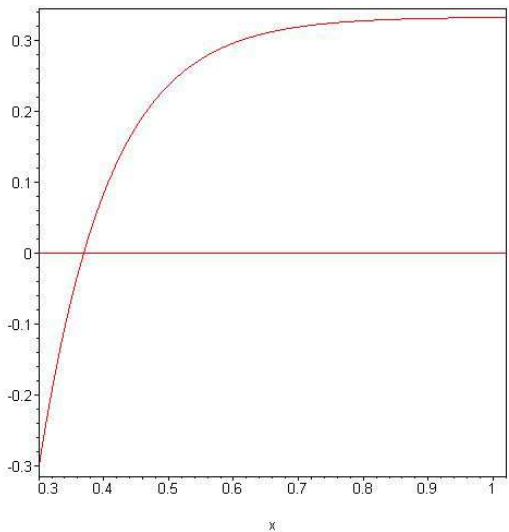}\\
%\epsfbox{fig1.eps}\\
\end{center}
\vskip -0.5cm \caption[fig1] {$\ddot{a}/(\Lambda a)$ in terms of
$x=t /t_{*}$ where $t_{*}$ is the age of our universe.} \label{fig1}
\end{figure}
%%%%%%%%%%%%%%%%%%%%%%%%%%%%%%%%%%%%%%%%%%%%%%%%%%%%%%%%%%%%%%%%%%%%%%%%%%%%%

%%%%%%%%%%%%%%%%%%%%%%%%%%%%%%%%%%%%%%%%%%%%%%%%%%%%%%%%%%%%%%%%%%%%%%%%%%%%%
\begin{figure}
\begin{center}
\includegraphics[width=7cm]{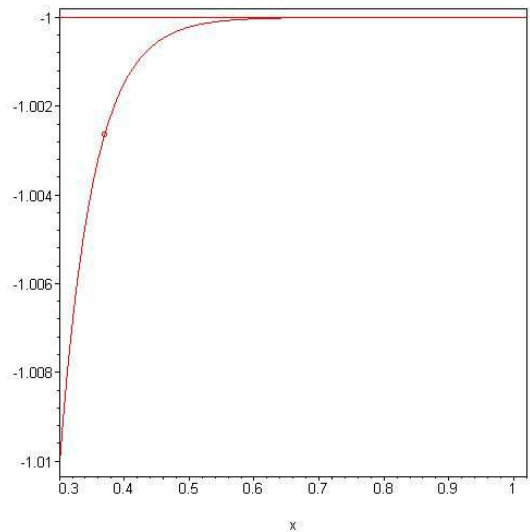}\\
%\epsfbox{fig1.eps}\\
\end{center}
\vskip -0.5cm \caption[fig2] {$\omega$ in terms of $x=t /t_{*}$.
The circle denotes the transition point from deceleration to
acceleration at $t_{tr}=0.37 t_{*}$.} \label{fig2}
\end{figure}
%%%%%%%%%%%%%%%%%%%%%%%%%%%%%%%%%%%%%%%%%%%%%%%%%%%%%%%%%%%%%%%%%%%%%%%%%%%%%

We depict in Fig 1 the graph of $\ddot{a}/(\Lambda a)$ and in Fig
2 the graph of $\omega$ in (\ref{omega2}) in term of the
dimensionless variable $x$. The circle in Fig 2 denotes the
transition point from deceleration to acceleration at $t_{tr}=0.37
t_{*}$ to yield the value of $\omega$ at that moment, \beq
\omega=\frac{79-21\sqrt{5}}{15-21\sqrt{5}}=-1.0026,\eeq which is
not ruled out by the current observation~\cite{phantom}. We assume
that our solution describes the recent acceleration of our
universe starting at time $t=t_{tr}$ and $b_{*}$ is very small
compared to $a_{*}$ so that the exponential increase of the size
of the extra dimensions does not contradict with the experimental
observations.

In summary, we found that ten-dimensional phantom model provides a
new cosmological solution of the type $a(t)\sim e^{\lambda(t)}$ with
$\lambda(t)=\frac{1}{6}\sqrt{\bar{\Lambda}}t-\frac{4n}{9}(e^{-\frac{3}{2}\sqrt{\bar{\Lambda}}t}-1).$
This function has the feature that it has a vanishing second time
derivative at time $t_{tr}=5.04$ Gyr, at which the recent
acceleration of the universe starts. This leads to the interesting
possibility that this model can give one way of understanding the
coincidence problem \cite{coinci}.

A couple of comments are in order. In the standard cosmology with a
cosmological constant $\Lambda$, there is a transition from the
matter-dominated era to $\Lambda$-dominated era.  It is shown that
there exists the nonsmooth curvature associated with the multiple
discontinuities at the transition~\cite{hong}. In our case with a
phantom field in higher dimensional gravity theory, the solution
belongs to the class of generalized exponential acceleration where
$a(t)\sim e^{F(t)}$ with $\ddot{a}(t)=0$ for some finite time $t$.
In this case, the cosmic transition occurs continuously at time $t=
t_{tr}$. This kind of approach %also
appeared in applying intermediate inflation models~\cite{interm} to
the late time acceleration with $F=At^f$ where $A>0$ and $0<f<1$. It
has the transition time given by
$t=\left(\frac{1-f}{Af}\right)^{1/f}$. This also appears in a scalar
phantom-non-phantom transition model \cite{nojiri} to unify phantom
inflation with late-time acceleration. We have neglected the matter
contribution in our analysis. Since $\ddot{a}$ is always negative
for matter contribution and it is negligible after $t=t_{tr}$, the
inclusion of matter part in our model will not change the
characteristic feature of the accelerating universe except that it
could shift the transition time slightly.

Our solution is obtained in ten-dimensional spacetime and crucially
depends on the condition (\ref{condi}). It would be interesting to
check whether relaxing these conditions could lead to some more
general solutions.

\acknowledgments  We would like to thank an anonymous referee for
helpful comments. The work of STH was supported by the Korea
Research Foundation (MOEHRD), Grant No. KRF-2006-331-C00071, and by
the Korea Research Council of Fundamental Science and Technology
(KRCF), Grant No. C-RESEARCH-2006-11-NIMS. THL was supported by the
Soongsil University Research Fund. PO was supported by the Science
Research Center Program of the Korea Science and Engineering
Foundation through the Center for Quantum Spacetime(CQUeST) of
Sogang University with grant number R11-2005-021.

\end{document}